\documentclass[twocolumn,showpacs,preprintnumbers,amsmath,amssymb]
{revtex4}

\usepackage{graphicx}
\usepackage{dcolumn}
\usepackage{bm}

\begin{document}

\title{Fundamental decoherence in quantum gravity}

\author{Rodolfo Gambini}
\affiliation{Instituto de Instituto de F\'{\i}sica, Facultad de
Ciencias, Igu\'a 4225,  Montevideo 11400, Uruguay.}
\author{Rafael A. Porto}
\affiliation{Department of Physics, Carnegie Mellon University,
Pittsburgh, PA 15213, USA}
\author{Jorge Pullin}
\affiliation{Department of Physics and Astronomy, Louisiana State
University, Baton Rouge, LA 70803-4001, USA}

\date{December 8th 2004}

\begin{abstract}
A recently introduced discrete formalism allows to solve the problem
of time in quantum gravity in a relational manner.  Quantum mechanics
formulated with a relational time is not exactly unitary and implies a
fundamental mechanism for decoherence of quantum states. The mechanism
is strong enough to render the black hole information puzzle
unobservable.
\end{abstract}
\maketitle

\section{Introduction}
The ``problem of time'' in quantum gravity arises largely due to the
presence of constraints in the theory, in particular the Hamiltonian
constraint (see \cite{Kuchar} for a review). If one could eliminate
the constraints almost all the conceptual problems with the problem of
time can be eliminated. In particular, one can implement the
quantization proposed by Page and Wootters \cite{PaWo} in which one
promotes all variables to quantum operators and chooses one of these
quantum variables to be the ``clock''. One then computes conditional
probabilities for other variables to take certain values when the
``clock'' variable is at a certain ``time''.

We have recently introduced \cite{Discrete} an approach to general
relativity that eliminates the constraints. The main idea is to
approximate the theory by a discrete theory, much like it is done in
particle physics when one approximates a field theory on the
lattice. The novelty of our approach is that the discrete theories we
construct are constraint-free, yet they can approximate general
relativity.  As a consequence, one can complete the Page--Wootters
quantization of the discrete theories and introduce a relational time
\cite{greece}.

An immediate consequence of having a ``quantum clock'' variable in
quantum mechanics is that the evolution is not unitary \cite{cqgdeco,
njp}. Both the clock and the system under study evolve unitarily and
under the usual rules of quantum mechanics in terms of a fiducial
background time $n$ (we use the letter $n$ to emphasize that we are
working in a discrete formulation, though this is not central to the
points discussed in this paper). This time is an inaccessible
variable, we could only measure it if we had a perfectly classical
clock.  What we can measure are the dynamical variables of the
problem, in particular $t$, the variable that describes the
clock. This variable is represented by a quantum operator and it will
have an expectation value and a dispersion. Upon evolution, the
dispersion will increase. One can show that if one prepares the clock
initially in a state in which $t$ is highly peaked around a given
value of the fiducial time $n$, the quantities under study (let us
call them $O$) will evolve according to an approximate Heisenberg
equation, but there will be corrective terms that imply that pure
states evolve into mixed states.

This kind of discussion is relevant to the black hole information
puzzle \cite{puzzle}. The puzzle arises because one could have a pure
state that undergoes gravitational collapse to form a black hole.  The
black hole will eventually evaporate leaving behind outgoing thermal
Hawking radiation. It will therefore be in a mixed state, so somehow
the initial pure state evolved into a mixed state. This is a problem
in ordinary quantum mechanics.  In quantum mechanics with a relational
time, since pure states decohere anyway, one should evaluate if the
rate of fundamental decoherence is slower or faster than the process
of black hole evaporation. If the state would have become totally
mixed anyway due to fundamental decoherence by the time the black hole
evaporates, then the puzzle is unobservable.

We have carried out some preliminary calculations to yield an estimate
of the fundamental decoherence and compare it to the black hole
evaporation rate \cite{bhinfo}. The rate of decoherence is related to
``how classical'' the clock one uses is.  Therefore we addressed the
question of what is the optimal clock that one can
construct. Following arguments of Salecker and Wigner and further
discussions by Ng and Amelino-Camelia \cite{salecker}, we established
that the optimal clock one can find is a black hole. The accuracy with
which a black hole can measure time is given by the frequency of its
(quasi)normal modes, which scales as the inverse of the black hole
mass. One would like therefore to have a black hole of small mass as a
clock in order to have more accuracy. But there is a limit to this, if
the mass is too small, the black hole-clock will evaporate too quickly
for one to observe the physics of interest. Therefore one has an
optimal accuracy one can achieve given a lapse of time that one wishes
to measure. This limit on the accuracy is clearly only theoretical, in
practice there will be other environmental factors that will affect
the accuracy of the clock. Even if one isolates the system from the
environment, there are quantum uncertainties (for instance, in the
position of the clock) that need to be taken into account. Here we
will only concentrate on the above mentioned effect, since it is one
of the effects of fundamental nature and can be viewed as an ultimate
limit on the accuracy of a clock.

The organization of this paper is as follows. In section II and III we
review results (\cite{njp,bhinfo}) on the evolution of conditional
probabilities and the estimate of an optimal clock, respectively. In
section IV we present a novel way of calculating explicitly the
evolution of the conditional probability in closed form for the black
hole information puzzle scenario.  This generalizes previous work in
which calculations were carried out to first non-trivial order in an
expansion \cite{bhinfo}.

\section{Evolution of the conditional probability}

To give a more quantitative version of things, let us compute
the time evolution of the conditional probability of measuring
an observable $O$ as a function of a real clock $t$ (for
further detail see \cite{njp}).

We start by considering the conditional probability defined we
mentioned above for an observable $O$ to take a certain value $o$
when the clock variable takes a value $t$,
\begin{equation}
{\cal P}(o \in \Delta o | t \in \Delta t)_\rho =
{ \sum_{n} {\rm Tr}\left( P_o(n) P_t(n) \rho P_t(n)\right) \over
\sum_n {\rm Tr}\left(P_t(n) \rho\right)},
\end{equation}
and one could have omitted the last $P_t(n)$ using the cyclicity of
the trace since we are assuming that $P_t$ and $P_o$ commute.
(If the observables ($O$ and $t$) have continuous spectra,
the projectors in the above expression should be understood as
integrated over the interval, i.e. $P_o(n) = \int_{\Delta o} P_{o'}(n)
do'$ and similar for $P_t$.) The sum over all possible fiducial times
$n$ is due to the fact that we do not know for which value of $n$ the
variable $t$ takes the value we want, and as the clock disperses there
will be several values of $n$ that correspond to $t$. We now introduce
the hypothesis that the clock and the rest of the system interact
weakly and write explicitly the evolution of the projectors in the
step parameter $n$ to get,
\begin{eqnarray}
&&{\cal P}(o \in \Delta o | t \in \Delta t)_\rho =\label{condprob}\\
&&= {
\sum_n {\rm Tr}\left(U^\dagger_2(n) P_o(0) U_2(n) U^\dagger_1(n) P_t(0) U_1(n) \rho_1 \otimes \rho_2\right)
\over \sum_n {\rm Tr}\left(P_t(n) \rho_1\right) {\rm Tr}\left(\rho_2\right)}\nonumber\\
&=& {
\sum_n {\rm Tr}\left(U^\dagger_2(n) P_o(0) U_2(n)\rho_2\right){\rm Tr}\left(
U^\dagger_1(n) P_t(0) U_1(n) \rho_1 \right)
\over \sum_n {\rm Tr}\left(P_t(n) \rho_1\right) {\rm Tr}\left(\rho_2\right)}.\nonumber
\end{eqnarray}
Here we have assumed the system breaks into two subsystems, the clock
(system 1) and the variables under study (system 2) and that the
density matrix for the total system is a direct product $\rho
=\rho_1\times \rho_2$ and $\rho_{1,2}$ are evolved in the fiducial
time $n$ by unitary evolution operators $U_{1,2}$.

{}From this expression, using the cyclic property of the trace, we can
identify the expressions of the density matrix evolved in relational
time. We start by defining the probability that the measurement $t$
corresponds to the value $n$,
\begin{equation}
{\cal P}_n(t) \equiv {{\rm Tr}\left(
P_t(0) U_1(n) \rho_1 U^\dagger_1(n) \right)
\over \sum_n
{\rm Tr}\left(
P_t(n)\rho_1  \right)},
\end{equation}
and notice that $\sum_n {\cal P}_n(t)=1$.

We now define the evolution of the density matrix,
\begin{eqnarray}
\tilde{\rho}_2(t)\equiv \sum_n U_2(n) \rho_2 U^\dagger_2(n) {\cal
P}_n(t), \label{evo1}
\end{eqnarray}
and noting that
\begin{equation}
{\rm Tr}\left(\tilde{\rho}_2(t)\right)
= \sum_n {\cal P}_n(t) {\rm Tr}\left(\rho_2\right)={\rm Tr}(\rho_2)
\end{equation}
one can equate the conditional probability (\ref{condprob}) with the
usual expression for a probability in quantum mechanics,
\begin{equation}
{\cal P}(o|t)_\rho \equiv
\frac{{\rm
Tr}(P_{o}(0)\tilde{\rho}(t))}{{\rm Tr}(\tilde{\rho}(t))},\label{usual}
\end{equation}
where the projector is evaluated at $t=0$ since in the Schr\"odinger
representation the operators do not evolve. From here on we drop
the subscript $2$ on the density matrix, since it is understood that
we are discussing the density matrix of the system under study.

It should be noted that all the sums in $n$, due to the assumption
that the time variable is semiclassical, are only nontrivial in the
small interval $\Delta_t n$ since outside of it, probabilities
vanish. Something else to notice is that when we introduced the
projectors, there was an integral over an interval.  Therefore in the
above expression for the evolution of the density matrix, this has to
be taken into account. Since the interval $\Delta t$ is arbitrary, one
can consider the limit in which its width tends to zero, apply the
mean value theorem in the integrals, and the interval in the numerator
and denominator cancel out, yielding an expression for
$\tilde{\rho}(t)$ that is independent of the interval, and involves
the non-integrated projector $P_t(0)$.

We have therefore ended with the standard probability expression with
an ``effective'' density matrix in the Schr\"odinger picture given by
$\tilde{\rho}(t)$.  In its definition, it is evident that unitarity
is lost, since one ends up with a statistical mixture of states
associated with different $n$'s.  We also notice that probabilities
are conserved, as can be seen by taking (\ref{usual}) and integrating
over $x$. We recall that $\tilde{\rho}$ is not the normalized
density matrix; the latter can be easily recovered dividing by the
trace.

We will assume that ${\cal P}_n(t) \equiv f(t-t_{max}(n))$ with
$f$ a function that decays quite rapidly for values of $t$ distant of
the maximum $t_{\rm max}$ which depends on $n$.

To manipulate expression (\ref{evo1}) more clearly, we will assume we
are considering a finite region of evolution and we are in the limit
in which the number of steps in that region is very large. We denote
the interval in the step variable $n$ as going from zero to $N$ with
$N$ a very large number. We define a new variable $v=\epsilon n$ with
dimensions of time such that $N \epsilon=V$ with $V$ a chosen finite
value. We can then approximate expression (\ref{evo1}) by a continuous
expression,
\begin{equation}
\tilde{\rho}(t)= \int_0^V dv f(t-t_{max}(v)) \rho(v).
\label{evocont}
\end{equation}

In this expression $t_{\rm max}(v)\equiv t_{\rm max}(n=v/\epsilon)$
and
\begin{equation}
\rho(v) = U_2(n=v/\epsilon) \rho U^\dagger_2(n=v/\epsilon).
\end{equation}
In all the above expressions, when we equate $n=v/\epsilon$ it should
be understood as $n={\rm Int}(v/\epsilon)$, which coincide in the
continuum limit. (Notice that strictly speaking we should write
$\rho(v/\epsilon)$ to keep the same functional form as for $\rho(n)$,
but we will drop the $\epsilon$ to simplify the notation.)
To simplify things further we will assume that we chose a physical
variable as our clock that has a linear relation with $v$, i.e.
$t_{\rm max}(v)\sim v$. In practice this is really not possible,
there will be departures from this linearity and this is another
effect that should be taken into account and will probably
lead to further decoherence.

\section{An optimally classical clock}

We now need to make some assumptions about the clock. As we argued
above, we use the ideas of Salecker--Wigner and Ng-Van Dam and
Amelino-Camelia \cite{salecker} to suggest that the ``most
classical'' clock one can build is a black hole. Briefly
described, the argument goes as follows: if one considers an ideal
isolated clock, it will lose accuracy as its wavefunction spreads.
To try to diminish this spread, one can increase the mass of the
clock. This process cannot continue indefinitely because
eventually one will have enough mass to produce a black hole
(trying to keep the density low by making a larger clock does not
work since then one has to take into account that matter is
elastic, etc). The black hole is therefore the most accurate clock
from this point of view and it is also attractive as a fundamental
clock given its fundamental nature (it is made of spacetime
itself). The accuracy of such a clock is given by the quasinormal
frequencies of the black hole, which scale as the inverse of the
mass. That is, to have a more accurate black hole clock, one needs
it to have a small mass. But a black hole of small mass evaporates
due to Hawking radiation quickly. This creates a tension between
these two requirements that leads to a formula that determines the
best accuracy one can achieve in the measurement of a time $T_{\rm
max}$,
\begin{equation}
\delta t \sim t_P \sqrt[3]{T_{\rm max}/t_P} \label{evap}
\end{equation}
where $t_P$ is Planck's time and from now on we choose units where
$\hbar=c=1$.

The reader might question why the spread of the wavefunction of
the clock limits its accuracy. After all, presumably the clock is
interacting with an environment which prevents the wavefunction
from spreading. We ignore this effect, since this interaction is
further source of inaccuracy in the clock (effects like this
have been studied in \cite{Garay}) and wish to concentrate
on the spread, which is an effect of fundamental nature, unrelated
to the environment.

\section{Application to the black hole information puzzle}

We need to make a quantum model of the black hole in order to study
its decoherence. Here we will make a very primitive model. We assume
the black hole horizon's area (or equivalently its energy) is
quantized. This is usually assumed in quantum black hole studies and
in particular it is predicted by loop quantum gravity. We choose a
basis of states for the black hole labeled by the energy (area).
The problem has some resemblance to the problem of an atom that
is in an excited state and emits radiation to reach its fundamental
state. If one considers the physical system under study to be the
atom plus the radiation field, its evolution is unitary. One would
expect a similar situation to hold for the black hole interacting
with the gravitational and matter fields surrounding it. Here is
where the paradox lies, since the evaporation process leads to loss
of unitarity for the total system. Our model will include information
about the black hole and the surrounding fields such that it
starts its evolution in a pure state, and we will study its
evolution according to the formalism developed in section II.
We consider
the system as described by a density matrix,
\begin{equation}
\rho = \sum_{ab} \rho_{ab} |\psi_a(t)><\psi_b(t)|,
\end{equation}
where
\begin{equation}
|\psi_a(t)>=|E(t)+\epsilon_a,E_0-E(t)>
\end{equation}
and where the first entry in the bra (ket) represents the energy
of the black hole at instant $t$, which changes with time in an
adiabatic fashion, the constant $E_0$ represents
the mean value of the
total energy of the system (which is conserved) and
$E_0-E(t)$ is the energy of the field at instant $t$. We consider
the state to be a superposition of states of the black hole that
differ in energy from $E(t)$ by $\epsilon_{a}$. To simplify the
analysis we consider only a pair of levels of energy that are
separated by an energy proportional to the temperature,
as one would expect for an evaporating hole.
Concretely, the
characteristic frequency for this energy is given by
\begin{equation}
\omega_{12}(t) = {1 \over \left(8 \pi\right)^2
t_P }\left(t_P\over T_{\rm max}-t\right)^{1/3}
\end{equation}
with $T_{\rm max}$ the lifetime of the black hole (how long it takes to
evaporate) and the subscript $12$ denotes that it is the transition
frequency between the two states of the system.  Although this model
sounds simple-minded it just underlies the robustness of the
calculation: it just needs that the black hole have discrete energy
levels characterized by a separation determined by the temperature of
the black hole.  It is general enough to be implemented either
assuming the Bekenstein spectrum of area or the spectrum stemming
from loop quantum gravity \cite{BaCaRo}.
We assume that we start with the black hole in a pure
state which is a superposition of different energy eigenstates
(there is no reason to assume that the black hole is  exactly in
an energy eigenstate, which would imply a stationary state with no
radiation being emitted; as soon as one takes into account the
broadening of lines due to interaction one has to consider a
superposition of states within the same broadened level with a
time dependent separation with a similar behavior).
Therefore the density matrix has off-diagonal
elements.

We now compute the evolution of the two level model for the
black hole using the formulas we developed in section II.
We consider the off-diagonal matrix element of the density
matrix in an energy eigen-basis, $\rho_{12}$. Its time
evolution is given by,
\begin{equation}
\rho_{12}(t) = \rho_{12}(0) \int_0^V dv{\cal P}_v(t)
\exp\left(i \int_0^v \omega_{12}(T) dT\right).
\end{equation}

We can now compute the integral in the exponent,
\begin{eqnarray}
\varphi_{12}&=&
\int_0^v \omega_{12}(T) dT = -{3 \over 2\left(8\pi\right)^2 t_P}
\times \\
&&\left\{\left[t_P \left(T_{\rm max}-v\right)^2\right]^{1/3}
-\left[t_P T_{\rm max}^2\right]^{1/3}\right\}.\nonumber
\end{eqnarray}

To compute the evolution we need to provide a model for ${\cal
P}_v(t)$.  We will assume it takes the form,
\begin{equation}
{\cal P}_v(t) = \Theta_{\tau(t)}(v-t) {1 \over \tau(t)}
\end{equation}
where the function $\Theta$ is one if $|v-t|<\tau/2$ and zero
otherwise, that is, a step function of width $\tau$ centered at
$t$. As we shall see the determination of the decoherence of the
state does not depend on the particular form of the width as a
function of $t$, one only needs to recall that the final width is
given by the limit for the accuracy of the clock computed in
equation (\ref{evap}).
\begin{equation}
\rho_{12}(t)={\rho_{12}(0)
 \over \tau(t)} \int_{t-\tau(t)/2}^{t+\tau(t)/2}
dv e^{i \varphi_{12}}\label{eq}.
\end{equation}
To compute the integral, and evaluate it for the value at
evaporation time $T_{\rm max}$ we make the variable transformation
$(T_{\rm max}-v )/t_P=u^3$, and write,
\begin{eqnarray}
\rho_{12}(T_{\rm max})&=& {\rho_{12}(0)
 \over \tau(T_{\rm max})} \exp\left(
{3 i \over 2 \left(8\pi\right)^2}
\left({T_{\rm max} \over t_P}\right)^{2/3}\right)\times\nonumber\\
&&
\int_{-U}^{U}
du u^2 t_P \exp\left({3 i u^2 \over 2 \left(8\pi\right)^2}\right)
\end{eqnarray}
with limits of integration $U={\sqrt[3]{-\tau(T_{\rm max})/ (2
t_P)}}$. For a Solar sized black hole $T_{\rm max}/t_P =
\left(M_{\rm Sun}/M_P\right)^3$, and therefore the integration
limits are large. The integral can be evaluated in closed form in
terms of Fresnel integrals, but it is more instructive to write
the asymptotic form. The modulus of the integral behaves
asymptotically as $\sqrt[3]{M_{\rm Sun}/M_P}$. One therefore has
an estimate for the modulus of the density matrix element
behavior,
\begin{equation}
|\rho_{12}(T_{\rm max})| \sim |\rho_{12}(0)| \left({M_P \over M_{\rm
 Sun}}\right)^{2 \over 3} \sim 10^{-28} |\rho_{12}(0)|.
\end{equation}
So for astrophysical black holes the puzzle is unobservable.
One could still ask what is the situation for black holes that
are smaller. We should recall that we have neglected several
effects that further imply decoherence, so it is likely that
the effect is larger than the estimate we present here.

\section{Discussion}

The calculation we have carried out here differs from those of our
previous papers. In \cite{bhinfo1} we made a first estimate of the
decoherence in the context of the black hole information puzzle.
In that first estimate we did not use an optimal clock and used a
cruder model of the spectrum of the black hole (temperature
independent). This calculation yielded that the quantum state did
not decohere entirely by the time of evaporation, though it
decohered in a significant amount. In \cite{bhinfo} we used an
optimal clock and an improved model of the spectrum of the hole,
but used only the first order expansion of the evolution equation
for the state as valid throughout the whole evolution, namely,
\begin{equation}
\frac{\partial\rho(t)}{\partial
t}=-i[H,\rho]-\sigma(t)[H,[H,\rho]],\label{doshs}
\end{equation}
where $\sigma(t)$ encodes the information about the quantum
fluctuations of the clock and it is related to the width $\tau(t)$
by $\sigma(t)\equiv \frac{d}{dt}\tau^2/24$. In \cite{bhinfo} we
did an explicit assumption for the form of $\tau(t)$, defining
$\tau^2(t)$ as the difference for the spread of the clock in the
interval $[0,T_{\rm max}]$ minus the spread in the interval
$[t,T_{\rm max}]$, that is

\begin{equation}
\tau^2(t)=t_P^2\left[\left(\frac{T_{\rm
max}}{t_P}\right)^{2/3}-\left(\frac{T_{\rm
max}-t}{t_P}\right)^{2/3}\right]\label{new},
\end{equation}

One can then determine $\sigma(t)$ in the expansion to be,

\begin{equation}
\sigma(t)=t_P/36 \left(\frac{t_P}{T_{\rm max}-t}\right)^{1/3}.
\end{equation}

The calculation in \cite{bhinfo} yielded the remarkable result
that it erased completely the information by the time evaporation
occurs. In this paper we have integrated the full evolution
equation and again one finds that there is a large level of
decoherence by the time of evaporation.\\
In the case of laboratory-like experiences our previous approach
is justified as roughly an expansion in $w_{12}\tau(t)$, where
$w_{12}$ is taken as a the natural energy gap of the system. In
order to see this we can integrate now (\ref{eq}) for a time
independent spectrum obtaining,
\begin{equation}
\rho_{12}(t)=2\frac{\rho_{12}(0)}{\tau(t)}e^{i\omega_{12}t}
\frac{\sin(\omega_{12}\frac{\tau(t)}{2})}{\omega_{12}}
\end{equation}
Expanding now in $\omega_{12}\tau(t)$,
\begin{eqnarray}
|\rho_{12}(t)|&=&2|\rho_{12}(0)|
\frac{sin(\omega_{12}\frac{\tau(t)}{2})}{\omega_{12}\tau(t)}\\
&&\sim
|\rho_{12}(0)|(1-\frac{1}{24}\omega^2_{12}\tau^2(t)\ldots)\nonumber
\end{eqnarray}

Noticing that if one is using the optimal clock for a total
interval $T$, according to (\ref{evap}) $\tau(T)=t_p^{2/3}T^{1/3}$
and therefore,
\begin{equation}
\omega^2_{12}\tau^2(t)= t_p^{4/3}\omega^2_{12}T^{2/3},
\end{equation}
we can immediately compare with a similar calculation with a two
level system, up to first order, obtained by using (\ref{doshs})
in \cite{bhinfo}, where the level of fundamental decoherence
is\footnote{The result in \cite{bhinfo} was computed with an
estimated $\sigma(t) \sim \left(\frac{t_P}{T_{\rm
max}-t}\right)^{1/3}$. The factor $1/36$ included here comes from
explicitly computing $\sigma={\dot b}$, where $b=
\int_{t-\tau/2}^{t+\tau/2} dv \frac{v^2}{2} ({\cal P}_v(t)-
\delta(v-t))=\tau^2/24$. See reference \cite{njp} for details.},

\begin{equation}
\log\left(\rho_{12}(T)\over \rho_{12}(0)\right)=-\frac{1}{24}
t_P^{(4/3)} T^{(2/3)} \omega_{12}^2.
\end{equation}

Both expressions are in agreement and the formalism is consistent.
The effect is too small to be observed in the lab, unless one can
construct a system with a significant energy difference between
the two levels. The most promising candidate systems would be
given by systems of ``Schr\"odinger cat'' type. Bose--Einstein
condensates could in some future provide a system where the effect
could be close to observability \cite{cqgdeco,sija}. On the other
hand, one could design an experiment where the effect is
intentionally large, i.e. choosing a clock that does not behave
semi-classically, as a proof of principle.\\
In the case of a Black Hole, where the spectrum is explicitly time
dependent, the calculation in \cite{bhinfo} can be also seen as an
approximation of the exact result in the case of astrophysical
black holes. Even though strictly speaking it is not a valid
expansion for $t=T_{\rm max}$, one can show that it is a good
approximation to the exact result provided $\tau(t) << T_{\rm
max}-t$, region in which the phase in (\ref{eq}) can be expanded
in series. This condition translate into $t_{\rm max}<< T_{\rm
max}- t_p^{2/3}T_{\rm max}^{1/3}$ and therefore covers a good
portion of the
$[0, T_{\rm max}]$ interval for $M_p/M<<1$.\\

Several caveats are in order. To begin with, it is clear that we
have taken a very crude model for the black hole and a more
detailed calculation is needed before one can completely write off
the black hole information puzzle as an observable effect, but the
present calculation provides good hope that the problem can indeed
be solved. A realistic calculation seems somewhat beyond the state
of the art. For instance, it is clear that the calculation should
model quantum mechanically the black hole but also the fields it
interacts with in a detailed way in the context of a theory of
quantum gravity. There is also the issue that in these
calculations we are neglecting the elaborate space-time structure
of the black hole and we are treating it as a ``star that
disappears'' in the sense that it occupies a finite region of
space and time. This is in order to have a definite ``evaporation
time'' to use in the calculation. In reality, a black hole that
evaporates will imply a change in the causal structure of
space-time that is yet to be understood. Properties like
information loss should eventually be properly framed in a
space-time context. This paper can only be viewed as a further
step towards understanding how the imperfection in clocks can
yield to loss of coherence in quantum mechanics.

Summarizing, we have shown that unitarity in quantum mechanics
only holds when describing the theory in terms of a perfect idealized
clocks. If one uses realistic clocks loss of unitarity is introduced.
We have estimated a minimum level of loss of unitarity based on
constructing the most accurate clocks possible. The loss of unitarity
is universal, affecting all physical phenomena. We have shown that
although the effect is very small, it may be important enough to
avoid the black hole information puzzle.

\section{Acknowledgments}
This work was supported by grant NSF-PHY0244335 and funds from the
Horace Hearne Jr. Institute for Theoretical Physics. The work of
R.A.P. is supported in part by DOE contracts DOE-ER-03-40682-143
and DE-AC02-6CH03000.

\bibliography{apssamp}

\begin{thebibliography}{00}
\bibitem{Kuchar} K. Kucha\v{r}, ``Time and interpretations of quantum
gravity'', in ``Proceedings of the 4th Canadian conference on general
relativity and relativistic astrophysics'', G. Kunstatter, D. Vincent,
J. Williams (editors), World Scientific, Singapore (1992).
[available online at http://www.phys.lsu.edu/faculty/pullin/kvk.pdf].
\bibitem{PaWo} D.~N.~Page and W.~K.~Wootters, Phys.\ Rev.\ D {\bf
27}, 2885 (1983)
\bibitem{Discrete}
R.~Gambini and J.~Pullin,
Phys.\ Rev.\ Lett.\  {\bf 90}, 021301 (2003)
[arXiv:gr-qc/0206055];
C. Di Bartolo, R. Gambini, J. Pullin,
Classical and Quantum Gravity {\bf 19}, 5475 (2002).
\bibitem{greece} R. Gambini, R.A. Porto and J. Pullin,
In ``Recent developments in gravity,''
K. Kokkotas, N. Stergioulas, editors,
World Scientific, Singapore (2003) [arXiv:gr-qc/0302064].
\bibitem{cqgdeco}
R. Gambini, R. Porto, J. Pullin,
Class.\ Quant.\ Grav.\  {\bf 21}, L51 (2004)
[arXiv:gr-qc/0305098].
\bibitem{njp} R. Gambini, R. Porto, J. Pullin,
New J.\ Phys.\  {\bf 6}, 45 (2004)
[arXiv:gr-qc/0402118].
\bibitem{puzzle}
See for instance S. Giddings, L. Thorlacius, in
``Particle and nuclear astrophysics and cosmology in the next
millennium'', E. Kolb (editor), World Scientific, Singapore (1996)
[arXiv:astro-ph/9412046]; for more recent references see
S.~B.~Giddings and M.~Lippert, [arXiv:hep-th/0402073] and D.~Gottesman
and J.~Preskill, JHEP {\bf 0403}, 026 (2004) [arXiv:hep-th/0311269].
\bibitem{bhinfo}
R. Gambini, R. Porto, J. Pullin, Phys. Rev. Lett. {\bf 93}, 240401
(2004) [arXiv:hep-th/0406260].
\bibitem{salecker}
E. Wigner, Rev. Mod. Phys. {\bf 29}, 255 (1957);
H. Salecker, E. Wigner, Phys. Rev. {\bf 109}, 571 (1958);
G.~Amelino-Camelia,
Mod.\ Phys.\ Lett.\ A {\bf 9}, 3415 (1994)
[arXiv:gr-qc/9603014];
Y.~J.~Ng and H.~van Dam,
Annals N.\ Y.\ Acad.\ Sci.\  {\bf 755}, 579 (1995)
[arXiv:hep-th/9406110];
Mod.\ Phys.\ Lett.\ A {\bf 9}, 335 (1994).
\bibitem{Garay}  I. Egusquiza, L. Garay, J. Raya, Phys. Rev. {\bf
A59}, 3236 (1999).
[arXiv:quant-ph/9811009]
\bibitem{BaCaRo} For a discussion of both spectra see,
M.~Barreira, M.~Carfora and C.~Rovelli,
Gen.\ Rel.\ Grav.\  {\bf 28}, 1293 (1996)
[arXiv:gr-qc/9603064].
\bibitem{bhinfo1}
R. Gambini, R. Porto, J. Pullin
``No black hole information puzzle in a relational universe,''
[arXiv:hep-th/0405183] to appear in Int. J. Mod. Phys. D
\bibitem{sija}
C.~Simon and D.~Jaksch, Phys. Rev. {\bf A70}, 052104 (2004)
 [arXiv:quant-ph/0406007].
\end{thebibliography}

\end{document}